\documentclass[12pt]{article}
\usepackage{epic}
\def\option{\noindent}
\def\vs{\vspace}
\def\mod{{\rm \;mod\;}}
\def\gcd{{\rm  gcd}}
\def\lcm{{\rm  lcm}}
\def\e{{\rm e}}

\def\eg{{\it e.g.}}
\newcommand{\Z}{\mathsf{Z}\kern -5pt \mathsf{Z}}
\def\half{ {1\over 2} }
\def\smhalf{  {\textstyle{1\over 2} }}
\def\id{ 0 } 
\def\a{\alpha}
\def\b{\beta}

\def\Lam{\Lambda}
\def\rrho{\mu}
\def\lam{\lambda}
\def\sig{\sigma}
\def\om{\omega}
\def\omc{{\om_c}}

\def\Zop{ Z^{\rm open} }
\def\xNK{ {x_{N,K}} }

\def\xKN{ {x_{K,N}} }
\def\xoddnoddk{ {x_{2n+1,2k+1}} }
\def\xoddkoddn{ {x_{2k+1,2n+1}} }
\def\xoddnevenk{ {x_{2n+1,2k}} }
\def\xevenkoddn{ {x_{2k,2n+1}} }

\def\xevennevenk{ {x_{2n,2k}} }
\def\xevenkevenn{ {x_{2k,2n}} }
\def\xevennK{ {x_{2n,K}} }

\def\Pplus{ P_+^K }
\def\zeroth{zero${}^{\rm th}$ }
\def\except{{\rm \;for\;} N+K {\rm \; even~(except~for~}N=K=2^m)}
\def\Spr{{S'}}
\def\Kp{{K'}}
\def\dpr{ {\prime\prime} }
\def\cH{  {\cal H}  }
\def\Exp{  {\cal E}^{\om}  }
\def\Expc{  {\cal E}^{\omc}  }
\def\Bound{  {\cal B}^{\om}  }
\def\Boundomc{  {\cal B}^{\omc}  }
\def\ta{{\tilde{\a}}}
\def\tSpr{{\tS'}}

\def\tn{{\tilde{n}}}

\def\bM{{\bf{M}}}
\def\tbM{{\widetilde{\bM}} }
\def\tM{{\widetilde{M}}}
\def\bOm{{\bf{\Omega}}}
\def\tQ{{\tilde{Q}}}
\def\tS{{\tilde{S}}}
\def\tildet{{\tilde{t}}}

\def\tlam{{\tilde{\lambda}}}
\def\tLam{{\tilde{\Lambda}}}
\def\tmu{{\tilde{\mu}}}

\def\tpsi{{\tilde{\psi}}}
\def\tphi{{\tilde{\phi}}}
\def\tadp{ {\widetilde{ \a^\dpr }} }
\def\ha{{\hat{\a}}}
\def\hb{{\hat{\b}}}
\def\hmu{{\hat{\mu}}}
\def\hrho{{\hat{\rrho}}}
\def\bz{{\bar z}}

\def\bV{\overline{V}}
\def\bJ{\overline{J}}

\def\bI{\overline{I}}
\def\bJ{\overline{J}}
\def\g{g}
\def\gcup{ {\breve g} }
\def\suN{{{\rm su}(N)}}

\def\sothree{{{\rm so}(3)}}

\def\suevenk{{{\rm su}(2k)}}
\def\suoddn{{{\rm su}(2n+1)}}

\def\sooddn{{{\rm so}(2n+1)}}
\def\sooddk{{{\rm so}(2k+1)}}
\def\spn{{{\rm sp}(n)}}
\def\spk{{{\rm sp}(k)}}
\def\hgK{{\hat{g}_K}}   
\def\twhgK{{\hat{g}_K^\om}}   
   
\def\suNK{{\widehat{\rm su}(N)_K}}

\def\suKN{{\widehat{\rm su}(K)_N}}
\def\suoddK{{\widehat{\rm su}(2n+1)_K}}
\def\suevenK{{\widehat{\rm su}(2n)_K}}
\def\sutwotwoK{{\widehat{\rm su}(2)_{2\Kp}}}
\def\suoddnK{{\widehat{\rm su}(2n+1)_K}}
\def\suoddnoddk{{\widehat{\rm su}(2n+1)_{2k+1}}}
\def\suoddnevenk{{\widehat{\rm su}(2n+1)_{2k}}}
\def\suoddkoddn{{\widehat{\rm su}(2k+1)_{2n+1}}}
\def\suevennevenk{{\widehat{\rm su}(2n)_{2k}}}
\def\suevenkevenn{{\widehat{\rm su}(2k)_{2n}}}
\def\suevenkoddn{{\widehat{\rm su}(2k)_{2n+1}}}
\def\soNK{{\widehat{\rm so}(N)_K}}
\def\soKN{{\widehat{\rm so}(K)_N}}
\def\soK{{\widehat{\rm so}(2n+1)_{\Kp}}}
\def\sothreeK{{\widehat{\rm so}(3)_\Kp}}
\def\soKpo{{\widehat{\rm so}(2n+1)_{K+1}}}
\def\soKpt{{\widehat{\rm so}(2n+1)_{K+2}}}
\def\sotkpo{{\widehat{\rm so}(2n+1)_{2k+1}}}
\def\sotkpt{{\widehat{\rm so}(2n+1)_{2k+2}}}
\def\sotkpth{{\widehat{\rm so}(2n+1)_{2k+3}}}
\def\sotnpo{{\widehat{\rm so}(2k+1)_{2n+1}}}
\def\sotnpt{{\widehat{\rm so}(2k+1)_{2n+2}}}
\def\sotnpth{{\widehat{\rm so}(2k+1)_{2n+3}}}

\def\spnk{{\widehat{\rm sp}(n)_k}}
\def\spkn{{\widehat{\rm sp}(k)_n}}
\def\spnKpnmo{{\widehat{\rm sp}(n)_{K+n-1}}}
\def\spnKpn{{\widehat{\rm sp}(n)_{K+n}}}

\def\AnnoneK{  (A_{2n}^{(1)})_{K}}
\def\AnntwoK{  (A_{2n}^{(2)})_{K}}
\def\AnnmoneK{  (A_{2n-1}^{(1)})_{K}}
\def\AnnmtwoK{  (A_{2n-1}^{(2)})_{K}}

\def\DntwoK{  (D_{n+1}^{(2)})_K  }
\def\twbdy{  | B \rangle\!\rangle^\om }

\def\twishimu{  | \mu \rangle\!\rangle_I^\om }
\def\Ctwishimu{  | \mu \rangle\!\rangle_I^\omc }

\def\cardya{  | \a \rangle\!\rangle_C^\om }
\def\Ccardya{  | \a \rangle\!\rangle_C^\omc }

\def\nblama{ {n_{\b\lam}}^{\a}  }
\def\tnblama{{    \tn_{\hb\tlam}^{~~~\ha} }}
\def\sblama{    {s_{\b\lam}}^{\a}  }
\def\tsblama{{  \tilde{s}_{\hb\tlam}^{~~~\ha}  }}
\def\four{  {\vcenter  {\vbox  
              {\hrule height.4pt
               \hbox {\vrule width.4pt  height3pt  
                      \kern3pt 
                      \vrule width.4pt  height3pt 
                      \kern3pt
                      \vrule width.4pt  height3pt 
                      \kern3pt
                      \vrule width.4pt  height3pt 
                      \kern3pt
                      \vrule width.4pt height3pt}
               \hrule height.4pt}
                         }
              }
           }
\def\oneoneoneone{ 
              {\vcenter  {\vbox  
              {\hrule height.4pt
               \hbox {\vrule width.4pt  height3pt  
                      \kern3pt 
                      \vrule width.4pt  height3pt }
               \hrule height.4pt
               \hbox {\vrule width.4pt  height3pt  
                      \kern3pt 
                      \vrule width.4pt  height3pt }
               \hrule height.4pt
               \hbox {\vrule width.4pt  height3pt  
                      \kern3pt 
                      \vrule width.4pt  height3pt }
               \hrule height.4pt
               \hbox {\vrule width.4pt  height3pt  
                      \kern3pt 
                      \vrule width.4pt  height3pt }
               \hrule height.4pt}
                         }
              }
           }
\def\two{  {\vcenter  {\vbox  
              {\hrule height.4pt
               \hbox {\vrule width.4pt  height3pt  
                      \kern3pt 
                      \vrule width.4pt  height3pt 
                      \kern3pt
                      \vrule width.4pt height3pt}
               \hrule height.4pt}
                         }
              }
           }
\def\oneone{ 
              {\vcenter  {\vbox  
              {\hrule height.4pt
               \hbox {\vrule width.4pt  height3pt  
                      \kern3pt 
                      \vrule width.4pt  height3pt }
               \hrule height.4pt
               \hbox {\vrule width.4pt  height3pt  
                      \kern3pt 
                      \vrule width.4pt  height3pt }
               \hrule height.4pt}
                         }
              }
           }
\def\be{\begin{equation}}
\def\ee{\end{equation}}
\def\bea{\begin{eqnarray}}
\def\eea{\end{eqnarray}}

\def\theequation{\thesection.\arabic{equation}}

\begin{document}
\bibliographystyle{bst}

\begin{flushright}
{\tt hep-th/0606147}\\
BRX-TH-576\\
BOW-PH-137\\
\end{flushright}
\vspace{30mm}

\vspace*{.3in}

\begin{center}
{\Large\bf\sf  Twisted D-branes of the $\suNK$ WZW model \\
and level-rank duality}
\vskip 5mm Stephen G. Naculich\footnote{Research supported in part
by the NSF under grant PHY-0456944}$^{,a}$
and Howard J.  Schnitzer\footnote{Research supported in part 
by the DOE under grant DE--FG02--92ER40706\\
{\tt \phantom{aaa} schnitzr@brandeis.edu; naculich@bowdoin.edu}\\
}$^{,b}$

\end{center}

\begin{center}
$^{a}${\em Department of Physics\\
Bowdoin College, Brunswick, ME 04011}

\vspace{.2in}

$^{b}${\em Theoretical Physics Group\\
Martin Fisher School of Physics\\
Brandeis University, Waltham, MA 02454}
\end{center}
\vskip 2mm

\begin{abstract}
We analyze the level-rank duality of $\om_c$-twisted D-branes 
of $\suNK$ (when $N$ and $K>2$).
When $N$ or $K$ is even, the duality map involves
$\Z_2$-cominimal equivalence classes of twisted D-branes.
We prove the duality of the spectrum of an open string
stretched between $\omc$-twisted D-branes, 
and ascertain the relation between the charges
of level-rank-dual $\omc$-twisted D-branes. 
\end{abstract}

\vfil\break


\section{Introduction}
\setcounter{equation}{0}
\label{secintro}

Level-rank duality is a relationship between 
various quantities in bulk Wess-Zumino-Witten models
with classical Lie groups \cite{
Naculich:1990hg, 
Altschuler:1989nm, 
Mlawer:1990uv}.
It has recently been shown \cite{
Naculich:2005tn,
Naculich:2006mt} 
that level-rank duality also applies to untwisted 
and to certain twisted D-branes 
in the corresponding boundary WZW models \cite{
Klimcik:1996hp}-\cite{
Gaberdiel:2004yn}.  
(For a review of D-branes on group manifolds, 
see ref.~\cite{Schomerus:2002dc}.)
In this paper, we extend this work to include 
all charge-conjugation-twisted D-branes of the $\suNK$ WZW model. 

Untwisted (i.e., symmetry-preserving) D-branes of WZW models
are labelled by the integrable highest-weight 
representations $V_\lam$ of the affine Lie algebra.
For $\suNK$, 
these representations belong to cominimal equivalence classes 
generated by the $\Z_N$ simple current of the WZW model,
and therefore so do the untwisted D-branes of the model.
Level-rank duality is a one-to-one correspondence 
between cominimal equivalence classes (or simple-current orbits) 
of integrable representations of $\suNK$ and $\suKN$,
and therefore induces a map between 
cominimal equivalence classes 
of untwisted D-branes.

The spectrum of an open string stretched between 
D-branes labelled by $\a$ and $\b$
is specified by the coefficients of the partition function 
\be
\label{eq:openpartition}
\Zop_{\a\b} (\tau) = \sum_{\lam \in \Pplus} \nblama \chi_\lam (\tau) 
\ee
where $\chi_\lam (\tau)$
is the affine character of the integrable highest-weight 
representation $V_\lam$.
For untwisted D-branes, the coefficients $\nblama$ are equal to
the fusion coefficients of the bulk WZW theory \cite{Cardy:1989ir},
so the well-known level-rank duality of the fusion 
rules \cite{
Naculich:1990hg,
Altschuler:1989nm,
Mlawer:1990uv}
implies the duality of the open-string spectrum between untwisted branes.

Untwisted D-branes of $\suNK$ possess a conserved D0-brane charge 
belonging to $\Z_{\xNK}$: 
\be
Q_\lam = (\dim \lam)_\suN \quad \mod \xNK
\ee
where \cite{
Fredenhagen:2000ei,
Maldacena:2001xj} 
\be
\label{eq:xnk}
\xNK \equiv {  N+K \over \gcd \{ N+K, \lcm \{ 1, \ldots, N-1\} \}  }\,.
\ee
The charges of cominimally-equivalent untwisted D-branes are equal
up to sign \cite{Maldacena:2001xj}
\be
\label{eq:untwistedcomin}
Q_{\sig (\lam)} = (-1)^{N-1}  Q_\lam ~\mod \xNK
\ee
where $\sig$ is the $\Z_N$ simple current of $\suNK$.
It was shown in refs.~\cite{Naculich:2005tn,Naculich:2006mt} 
that the charges of level-rank-dual untwisted D-branes of $\suNK$ 
and $\suKN$ 
are related by
\be
\label{eq:untwistedduality}
\tQ_\tlam = \left\{ 
(-1)^{r(\lam)}  Q_\lam \quad \mod x
\qquad {\rm for~} N+K {\rm ~odd} 
\qquad\qquad\qquad\qquad\qquad\qquad
\atop
\quad\quad\quad Q_\lam \quad \mod  x
\qquad \except
\right. 
\ee
where $r(\lam)$ is the number of boxes in the Young tableau 
associated with the representation $\lam$,
where $\tlam$ is the level-rank-dual representation
of $\suKN$ 
associated with the transposed tableau,
and $x = \min \{ \xNK, \xKN  \}$.
For the remaining case, it was conjectured that
\be
\label{eq:specialduality}
\tQ_\tlam = \left\{ 
{ (-1)^{r(\lam)/N} Q_\lam  \quad  \mod x~~~ {\rm~when~}  N~|~r(\lam) \atop
\quad\quad\quad  \quad\quad         
Q_\lam  \quad  \mod x~~~ {\rm~when~}  N~\not |~r(\lam) }
\right\} \quad
{\rm~for~}  N=K=2^m
\ee
on the basis of numerical evidence.

In addition to untwisted D-branes,
most WZW models contain twisted D-branes, 
whose charges also belong to $\Z_{\xNK}$ \cite{
Bouwknegt:2000qt,
Freed:2001jd, 
Braun:2003rd, 
Gaberdiel:2004hs}.
The coefficients $\nblama$ of the partition function (\ref{eq:openpartition})
of an open string stretched between twisted D-branes $\a$ and $\b$
are given by 
\be
\label{eq:opencoeff}
\nblama =  \sum_{\mu \in \Exp}
{ \psi^*_{\a\mu} S_{\lam\mu} \psi_{\b\mu} \over S_{\id\mu} } 
\ee
where $\psi_{\a\mu}$ 
is the modular-transformation matrix 
of the associated twisted affine Lie algebra.

One such class of D-branes for $\suNK$ 
are those twisted by the charge-conjugation symmetry $\omc$,
which exist for all $N>2$.
This paper will analyze the level-rank duality 
of $\omc$-twisted D-branes of $\suNK$ (for $N$ and $K>2$),
and in particular, 
the relationship between the open-string partition function 
coefficients (\ref{eq:opencoeff}), 
and between the D-brane charges.
(In ref.~\cite{Naculich:2006mt}, 
level-rank duality of $\omc$-twisted
D-branes was examined in the special
case that $N$ and $K$ were both odd.)

As shown in ref.~\cite{Gaberdiel:2002qa},
and reviewed in sections \ref{secsueven} and \ref{secsuodd}, 
the $\omc$-twisted D-branes of $\suevenK$ (resp. $\suoddK$)
are labelled by a subset of integrable highest-weight 
representations of $\soKpo$ (resp. $\soKpt$),
or alternatively, by a subset of integrable highest-weight
representations of $\spnKpnmo$ (resp. $\spnKpn$).
In section \ref{secsueven}, 
we show that, like untwisted D-branes,
$\omc$-twisted D-branes of $\suevenK$
belong to cominimal equivalence classes, 
but now generated by the $\Z_2$ simple current of $\soKpo$.
As shown in section \ref{seccharge},
cominimally-equivalent $\omc$-twisted D-branes of $\suevenK$ 
have equal and opposite charges (mod $\xevennK$).

In section \ref{secduality}, 
we describe a one-to-one map $\a \to \ha$
between the $\omc$-twisted D-branes
(or cominimal equivalence classes of branes) of $\suNK$ and
the $\omc$-twisted D-branes
(or cominimal equivalence classes of branes) of $\suKN$.
The exact form of the level-rank map depends
on whether $N$ and $K$ are even or odd.
We then show the equality of the 
open string partition function coefficients
(\ref{eq:opencoeff}) for level-rank-dual $\omc$-twisted D-branes.
Because the level-rank map involves cominimal equivalence classes 
in the case of $\suevenK$, the natural quantity to consider in
that case is 
\be 
\label{eq:firstsblamadef}
\sblama 
= \left(\half\right)^{ \half [t(\a)+t(\b)] + 1 }
\left[ {n_{\b\lam}}^{\a}  + {n_{\b\lam}}^{\sig(\a)}  
+ {n_{\sig(\b)\lam}}^{\a}  + {n_{\sig(\b)\lam}}^{\sig(\a)}  \right]
\ee
where $\sig$ is the $\Z_2$ simple-current symmetry  of $\soKpo$,
and $t(\a)$ is defined in eq.~(\ref{eq:tdef}).

In section \ref{seccharge},
we ascertain the relationship between the charges 
of level-rank-dual $\omc$-twisted D-branes.

Sections \ref{sectwisted}
and \ref{secsoK} 
contain some necessary background material
on twisted states in WZW models and 
on integrable representations of $\soK$,
and concluding remarks comprise  section \ref{secconcl}.

\section{Twisted D-branes of WZW models} 
\setcounter{equation}{0}
\label{sectwisted}

In this section, we review some aspects of twisted D-branes of WZW models
and their relation to the twisted Cardy and twisted Ishibashi states 
of the closed-string sector, drawing on 
refs.~\cite{
Behrend:1998fd,
Fuchs:1999zi,
Birke:1999ik,
Ishikawa:2001zu,
Gaberdiel:2002qa}.

The WZW model,
which describes strings propagating on a group manifold,
is a rational conformal field theory
whose chiral algebra (for both left- and right-movers)
is the (untwisted) affine Lie algebra $\hgK$ at level $K$. 
We only consider WZW theories with a diagonal closed-string spectrum:
\be
\label{eq:diagonal}
\cH^{\rm closed}  
= \bigoplus_{\lam \in \Pplus}  
 V_\lam \otimes \bV_{\lam^*}
\ee
where $V$ and $\bV$ represent left- and right-moving states
respectively,
and $\lam^*$ denotes the representation conjugate to $\lam$.
$V_\lam \in \Pplus$ are integrable highest-weight representations of $\hgK$, 
whose highest weight $\lam$
has non-negative Dynkin indices $(a_0, a_1, \cdots, a_n)$ 
satisfying
$\sum_{i=0}^n  m_i a_i = K$
(where 
$n = {\rm rank~} \g$ and 
$(m_0, m_1, \cdots, m_n)$ are the dual Coxeter labels of $\hgK$).

D-branes of the WZW model may be studied algebraically 
in terms of the possible boundary conditions 
that can consistently be imposed on a WZW model with boundary.
We label the allowed boundary conditions 
(and therefore the D-branes) by $\a$, $\b$, $\cdots$.

We consider boundary conditions 
on the currents of the affine Lie algebra
of the form 
\be
\label{eq:twconditions}
\left[ J^a(z) - \omega \bJ^a(\bz)\right] \bigg|_{z=\bz} = 0
\ee
where $\omega$ is an automorphism of the Lie algebra $\g$.
These boundary conditions leave unbroken the $\hgK$ symmetry, 
as well as the conformal symmetry, of the theory.
Untwisted D-branes correspond to $\om = 1$.
Open-closed string duality allows one to correlate 
the boundary conditions (\ref{eq:twconditions}) 
of the boundary WZW model 
with coherent states $\twbdy \in \cH^{\rm closed}$ 
of the bulk WZW model satisfying
\be
\label{eq:twmodes}
\left[  J^a_m + \omega \bJ^a_{-m} \right] \twbdy = 0\,, \qquad m\in \Z
\ee
where $J^a_m$ are the modes of the affine Lie algebra generators.

Solutions of eq.~(\ref{eq:twmodes})
that belong to a single sector 
$ V_\mu \otimes \bV_{\om(\mu)^*} $
of the bulk WZW theory 
are known as $\om$-twisted Ishibashi states $\twishimu$.
(Solutions corresponding to $\om=1$ are the ordinary 
untwisted Ishibashi states \cite{Ishibashi:1988kg}.)
Since we are considering the diagonal closed-string theory
(\ref{eq:diagonal}),
these states only exist when $\mu = \om(\mu)$,
so the $\om$-twisted Ishibashi states are labelled by
$\mu \in \Exp$, where $\Exp \subset \Pplus$ are the integrable
highest-weight representations of $\hgK$ that satisfy $\om(\mu)=\mu$.
Equivalently, $\mu$ corresponds to an integrable highest-weight representation
of $\gcup$,  
the orbit Lie algebra \cite{Fuchs:1995zr} associated with $\hgK$.

A coherent state $\twbdy$ that corresponds to an
allowed boundary condition
must also satisfy additional (Cardy) conditions \cite{Cardy:1989ir}.
Solutions of eq.~(\ref{eq:twmodes}) that also satisfy the
Cardy conditions are denoted $\om$-twisted Cardy states $\cardya$,
where the labels $\a$ take values in some set $\Bound$.
The $\om$-twisted D-branes of $\hgK$ correspond to $\cardya$ 
and are therefore also labelled by $\a \in \Bound$.
These states correspond \cite{Birke:1999ik}
to integrable highest-weight representations 
of the $\om$-twisted affine Lie algebra $\twhgK$ 
(but see ref.~\cite{Alekseev:2002rj}).

The $\om$-twisted Cardy states may be expressed 
as linear combinations of $\om$-twisted Ishibashi states
\be
\label{eq:twcardyishi}
\cardya = \sum_{\mu \in \Exp}
{\psi_{\a\mu} \over \sqrt{S_{\id\mu}}} \twishimu 
\ee
where $S_{\lam\mu}$ is the modular transformation matrix of $\hgK$,
$\id$ denotes the identity representation,
and the coefficients $\psi_{\a\mu}$ may be identified \cite{Birke:1999ik}
with the modular transformation matrices of characters 
of the twisted affine Lie algebra $\twhgK$ \cite{Kacbuk},
as may be seen, for example,
by examining the partition function of an open string 
stretched between an $\om$-twisted and an 
untwisted D-brane \cite{Ishikawa:2001zu,Gaberdiel:2002qa}.
Using arguments presented, \eg, in ref.~\cite{Gaberdiel:2002qa},
the coefficients of the open string partition function
(\ref{eq:openpartition})
may be expressed as
\be
\label{eq:ndef}
\nblama =  \sum_{\mu \in \Exp}
{ \psi^*_{\a\mu} S_{\lam\mu} \psi_{\b\mu} \over S_{\id\mu} }  \,.
\ee

\section{Integrable representations of $\soK$}
\setcounter{equation}{0}
\label{secsoK}

This section presents details about integrable highest-weight
representations of $\soK$ that will be needed 
for the discussion of $\omc$-twisted states of the $\suNK$ WZW model.

Integrable representations of $\soK$ have Dynkin indices 
$(a_0, a_1, \cdots, a_{n})$
that satisfy $\sum_{i=0}^n m_i a_i = \Kp$,
where $m_i$ are the dual Coxeter labels of the 
extended Dynkin diagram for $\sooddn$

\begin{picture}(500,80)(10,10)
\put(100,40){\circle{5}}
\put(98,25){1}
\put(102,40){\line(1,0){26}}

\put(130,70){\circle{5}}
\put(134,66){1}
\put(130,68){\line(0,-1){26}}

\put(130,40){\circle{5}}
\put(128,25){2}
\put(132,40){\line(1,0){26}}

\put(160,40){\circle{5}}
\put(158,25){2}
\put(162,40){\line(1,0){26}}

\put(190,40){\circle{5}}
\put(188,25){2}
\put(192,40){\line(1,0){26}}

\put(220,40){\circle{5}}
\put(218,25){2}
\put(222,40){\line(1,0){26}}

\put(250,40){\circle{5}}
\put(248,25){2}
\put(252,40){\line(1,0){4}}
\put(258,40){\line(1,0){3}}
\put(263,40){\line(1,0){3}}
\put(268,40){\line(1,0){3}}
\put(273,40){\line(1,0){5}}

\put(280,40){\circle{5}}
\put(278,25){2}
\put(282,40){\line(1,0){26}}

\put(310,40){\circle{5}}
\put(308,25){2}
\put(312,41){\line(1,0){26}}
\drawline(327,40)(322,45)
\drawline(327,40)(322,35)
\put(312,39){\line(1,0){26}}

\put(340,40){\circle{5}}
\put(338,25){1}

\end{picture}
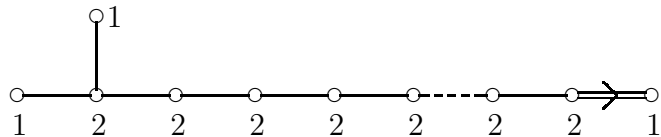
\option
(with the dual Coxeter labels shown adjacent to each node),
that is,\footnote{
\label{fnsothree}
\noindent
{\bf Note:} throughout this paper, by $\sothreeK$ we mean the affine
Lie algebra $\sutwotwoK$.
Its integrable representations have $\sothree$ Young tableaux 
that obey $\ell_1 \le \Kp$.
Since $\ell_1 = \half a_1$, this means that eq.~(\ref{eq:sonintegrable})
is replaced with $a_0 + a_1 = 2\Kp$ when $n=1$.}
\be
\label{eq:sonintegrable}
a_0 + a_1 + 2(a_2 + \cdots + a_{n-1}) + a_n = \Kp\,.
\ee
An even or odd value of $a_n$ corresponds
to a tensor or spinor representation respectively.
With each tensor representation of $\sooddn$ 
may be associated a Young tableau 
whose row lengths $\ell_i$ are given by
\be
\label{eq:sonrowlengths}
\ell_i = \left\{  
\smhalf a_n + \sum_{j=i}^{n-1} a_j \qquad \quad
{\rm for~} 1 \le i \le n-1 
\atop
~~\smhalf a_n \qquad\qquad\quad\qquad 
{\rm for~} i=n \qquad\qquad
\right.
\ee
with total number of boxes $r = \sum_{i=1}^n \ell_i$.
We also formally use eq.~(\ref{eq:sonrowlengths}) to define row lengths
for a spinor representation.
These row lengths are all half-integers,
and correspond to a ``Young tableau'' with a column of ``half-boxes.''
The integrability condition (\ref{eq:sonintegrable}) corresponds to
the constraint $\ell_1 + \ell_2 \le \Kp$ on the row lengths of the
tableau.

The extended Dynkin diagram of $\sooddn$ has a $\Z_2$ symmetry that interchanges
the $0^{\rm th}$ and $1^{\rm st}$ nodes.
This symmetry induces a simple-current symmetry 
(denoted by $\sig$)
of the $\soK$ WZW model
that pairs integrable representations
related by $a_0 \leftrightarrow a_1$,
with the other Dynkin indices unchanged.
Their respective Young tableaux are related by 
$\ell_1 \to \Kp - \ell_1$.
Under $\sig$, tensor representations are mapped to tensors, 
and spinor representations to spinors,
and the modular transformation matrix $S'$ of $\soK$ 
obeys \cite{Mlawer:1990uv}
\be
\label{eq:soncomin}
S'_{\sig(\a')\mu'}  = \pm S'_{\a'\mu'}  \quad 
{\rm for~} \mu'{\rm~a~}
\left\{ {\rm tensor} \atop {\rm spinor} \right\}  {\rm ~representation.} 
\ee
Representations related by $\sig \in \Z_2$ 
belong to a simple-current orbit, or cominimal equivalence class.

In this paper, we will refer to representations of $\soK$ with 
$\ell_1 < \smhalf \Kp$,
$\ell_1 = \smhalf \Kp$, and 
$\ell_1 > \smhalf \Kp$ 
as being of types I, II, and III respectively.
Type II representations are cominimally self-equivalent,
and are tensors (resp. spinors) when $\Kp$ is even (resp. odd).
Each simple-current orbit of $\soK$ contains either 
a type I and type III representation, 
or a single type II representation.

\section{Twisted states of the $\suevenK$ model} 
\setcounter{equation}{0}
\label{secsueven}

The invariance under reflection 
of the Dynkin diagram of the finite Lie algebra $\suN$ 
gives rise (when $N>2$) to 
an order-two automorphism $\omc$ of the Lie algebra,
under which the Dynkin indices $a_i$ ($i=1, \cdots, N-1$)
of an irreducible representation 
are mapped to $a_{N-i}$, corresponding to charge conjugation.
This automorphism lifts to an automorphism of the affine Lie algebra $\suNK$
that leaves the \zeroth node of the extended Dynkin diagram invariant. 
It gives rise (for $N>2$)
to a set of $\omc$-twisted Ishibashi states
and $\omc$-twisted Cardy states of the bulk $\suNK$ WZW model,
and a corresponding class of $\omc$-twisted D-branes of the boundary model.
In this section and the next, 
we review these twisted states for $\suevenK$ and $\suoddK$ respectively.
Much of this material is a summary of ref.~\cite{Gaberdiel:2002qa}.

\vs{.1in}
\noindent{\bf Twisted Ishibashi states of $\suevenK$}
\vs{.1in}

\option 
Recall from section \ref{sectwisted} 
that the $\omc$-twisted Ishibashi states $\Ctwishimu$ of 
the $\suevenK$ WZW model ($n>1$ is understood throughout this section)
are labelled by self-conjugate integrable highest-weight 
representations $\mu \in \Expc$ of $\suevenK = \AnnmoneK$.
These representations have Dynkin indices 
$(\mu_0, \mu_1, \cdots, \mu_{n-1}, \mu_n, \mu_{n-1}, \cdots, \mu_1)$
that satisfy 
\be
\label{eq:integrableIshieven}
\mu_0 + 2(\mu_1 + \cdots + \mu_{n-1}) + \mu_n  = K \,.
\ee
Equivalently, the $\omc$-twisted Ishibashi states of $\suevenK$ may be
characterized \cite{Fuchs:1995zr} by the  
integrable highest weight representations 
of the associated orbit Lie algebra $\gcup =\DntwoK$,
whose Dynkin diagram is

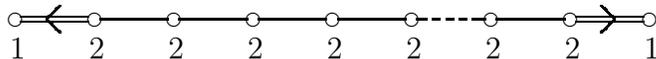
\begin{picture}(500,50)(10,10)
\put(100,40){\circle{5}}
\put(98,25){1}
\put(102,39){\line(1,0){26}}
\drawline(112,40)(117,45)
\drawline(112,40)(117,35)
\put(102,41){\line(1,0){26}}

\put(130,40){\circle{5}}
\put(128,25){2}
\put(132,40){\line(1,0){26}}

\put(160,40){\circle{5}}
\put(158,25){2}
\put(162,40){\line(1,0){26}}

\put(190,40){\circle{5}}
\put(188,25){2}
\put(192,40){\line(1,0){26}}

\put(220,40){\circle{5}}
\put(218,25){2}
\put(222,40){\line(1,0){26}}

\put(250,40){\circle{5}}
\put(248,25){2}
\put(252,40){\line(1,0){4}}
\put(258,40){\line(1,0){3}}
\put(263,40){\line(1,0){3}}
\put(268,40){\line(1,0){3}}
\put(273,40){\line(1,0){5}}

\put(280,40){\circle{5}}
\put(278,25){2}
\put(282,40){\line(1,0){26}}

\put(310,40){\circle{5}}
\put(308,25){2}
\put(312,41){\line(1,0){26}}
\drawline(327,40)(322,45)
\drawline(327,40)(322,35)
\put(312,39){\line(1,0){26}}

\put(340,40){\circle{5}}
\put(338,25){1}

\end{picture}

\option
with the integers adjacent to each node 
indicating the dual Coxeter label $m_i$.
The representation $\mu \in \Expc$ corresponds
to the $\DntwoK$ representation with Dynkin indices
$(\mu_0, \mu_1,  \cdots,  \mu_n)$,
whose integrability condition is precisely (\ref{eq:integrableIshieven}).

Each $\omc$-twisted Ishibashi state $\mu$ of $\suevenK$ 
may be mapped \cite{Gaberdiel:2002qa}
to an integrable highest-weight representation $\mu'$ 
of the untwisted affine Lie algebra $\soKpo$ 
with Dynkin indices
$(\mu_0 + \mu_1 +1, \mu_1, \cdots, \mu_n)$.
The constraint (\ref{eq:integrableIshieven}) translates into the
constraint $\ell_1 (\mu') \le \smhalf K$ on the $\soKpo$ Young tableaux.
This means that 
{\it $\omc$-twisted Ishibashi states of $\suevenK$
are in one-to-one correspondence with the
set of type I tensor and type I spinor representations of $\soKpo$.}

\vs{.2in}
\noindent{\bf Twisted Cardy states of $\suevenK$}
\vs{.1in}

\option 
Recall that the $\omc$-twisted Cardy states $\Ccardya$
(and therefore the $\omc$-twisted D-branes) 
of the $\suevenK$ WZW model
are labelled \cite{Birke:1999ik}
by the integrable highest-weight representations 
$\a \in \Boundomc$ 
of the twisted affine Lie algebra $\twhgK = \AnnmtwoK$,
whose Dynkin diagram is

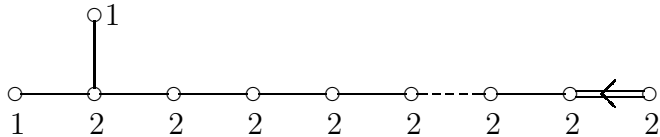
\begin{picture}(500,80)(10,10)
\put(100,40){\circle{5}}
\put(98,25){1}
\put(102,40){\line(1,0){26}}

\put(130,70){\circle{5}}
\put(134,66){1}
\put(130,68){\line(0,-1){26}}

\put(130,40){\circle{5}}
\put(128,25){2}
\put(132,40){\line(1,0){26}}

\put(160,40){\circle{5}}
\put(158,25){2}
\put(162,40){\line(1,0){26}}

\put(190,40){\circle{5}}
\put(188,25){2}
\put(192,40){\line(1,0){26}}

\put(220,40){\circle{5}}
\put(218,25){2}
\put(222,40){\line(1,0){26}}

\put(250,40){\circle{5}}
\put(248,25){2}
\put(252,40){\line(1,0){4}}
\put(258,40){\line(1,0){3}}
\put(263,40){\line(1,0){3}}
\put(268,40){\line(1,0){3}}
\put(273,40){\line(1,0){5}}

\put(280,40){\circle{5}}
\put(278,25){2}
\put(282,40){\line(1,0){26}}

\put(310,40){\circle{5}}
\put(308,25){2}
\put(312,41){\line(1,0){26}}
\drawline(322,40)(327,45)
\drawline(322,40)(327,35)
\put(312,39){\line(1,0){26}}

\put(340,40){\circle{5}}
\put(338,25){2}

\end{picture}

\option
The Dynkin indices 
$(a_0, a_1, \cdots, a_n)$ of the highest weights $\a$  
thus satisfy
\be
\label{eq:integrableCardyeven}
a_0 + a_1 + 2 (a_2 + \cdots + a_n) = K\,.
\ee
(For $n=2$, the twisted affine Lie algebra is instead
$D_3^{(2)}$ with nodes 1 and 2 interchanged \cite{Gaberdiel:2002qa},
but the condition (\ref{eq:integrableCardyeven}) remains valid.) 

The $\omc$-twisted Cardy state $\a \in \Boundomc$ of $\suevenK$
may be associated \cite{Gaberdiel:2002qa}
with an integrable highest-weight 
{\it spinor} representation $\a'$ of the untwisted affine Lie algebra $\soKpo$ 
with Dynkin indices 
$(a_0, a_1, \cdots, a_{n-1}, 2 a_n + 1 )$.
The constraint (\ref{eq:integrableCardyeven}) is precisely
the condition on integrable representations  of $\soKpo$.
(In terms of $\sooddn$ Young tableaux row lengths,
this constraint reads
$\ell_1 (\a') + \ell_2 (\a') \le  K+1$.)
Therefore, 
{\it there is a one-to-one correspondence between
the $\omc$-twisted D-branes of $\suevenK$
and integrable spinor representations of $\soKpo$ 
of type I, type II (when $K$ is even), and type III.}
For later convenience, we define
\be
\label{eq:tdef}
t(\a) = 
\left\{  
{0,  \quad \quad {\rm if~}\ell_1(\a') \neq \half (K+1)
 \qquad {\rm (types~I~and~III)}
\atop
1,  \quad \quad {\rm if~}\ell_1(\a') = \half (K+1)
 \qquad {\rm (type~II)\,.}
\qquad
\,\quad
}
\right.
\ee

Even though the $\omc$-twisted Cardy states 
and the $\omc$-twisted Ishibashi states of $\suevenK$
are characterized differently 
in terms of integrable representations of $\soKpo$,
they are equal in number. 
The $\omc$-twisted Cardy states $\a$  may be written 
as linear combinations of $\omc$-twisted Ishibashi states $\mu$,
with the transformation coefficients $\psi_{\a\mu}$ 
given by the modular transformation matrix of $\AnnmtwoK$.
In ref.~\cite{Gaberdiel:2002qa}, it was shown that, for $\suevenK$, 
these coefficients are proportional to matrix elements of the 
(real) modular transformation matrix  $S'$
of the untwisted affine Lie algebra $\soKpo$:
\be
\label{eq:evenpsi}
\psi_{\a\mu} =   \sqrt{2}  S'_{\a'\mu'}
             =   \sqrt{2}  S^{\prime*}_{\a'\mu'}
\ee
where $\a'$ and $\mu'$ are the $\soKpo$ representations related
to $\a$ and $\mu$ as described above.

Since the finite Lie algebra associated with
the twisted affine Lie algebra $\AnnmtwoK$ is $C_n$, 
the representations of $\AnnmtwoK$ form $C_n$-multiplets at each level.
More specifically \cite{Gaberdiel:2002qa},
each $\omc$-twisted Cardy state $\a \in \Boundomc$ of $\suevenK$
may be associated 
with an integrable highest-weight representation $\a^\dpr$ 
of the untwisted affine Lie algebra\footnote{Throughout this paper,
our convention is $\spn = C_n$.} $\spnKpnmo$
with (finite) Dynkin indices 
$(a_1, \cdots,  a_n)$.
The row lengths of 
the $\spn$ Young tableau associated with $\a^\dpr$ 
are equal to those of
the $\sooddn$ Young tableau associated with $\a'$ 
reduced by one-half: 
$\ell_i(\a'') = \ell_i(\a') - \half$.
Therefore,  
an alternative characterization of the 
$\omc$-twisted D-branes of $\suevenK$
is as the subset of integrable representations of $\spnKpnmo$
characterized by Young tableaux with row lengths satisfying
$\ell_1 (\a^\dpr) + \ell_2 (\a^\dpr) \le  K$.

\vs{.1in}
\noindent{\bf Equivalence classes of $\omc$-twisted D-branes of $\suevenK$}
\vs{.1in}

\option
The $\Z_2$ simple current symmetry $\sig$ of $\soKpo$ relates
type I and type III representations in pairs.
Using the 1-1 correspondence between 
integrable $\soKpo$ spinor representations
and $\omc$-twisted Cardy states, 
we lift the map $\sig$ to the twisted D-branes of $\suevenK$, 
and refer to $\sig(\a)$ as cominimally equivalent to $\a$.
(In section~\ref{seccharge}, we will show that $\a$ and $\sig(\a)$ 
have equal and opposite  D0-brane charges, modulo $\xevennK$.)
Therefore, {\it the cominimal equivalence classes of 
$\omc$-twisted D-branes of $\suevenK$ 
are in one-to-one correspondence with 
the set of type I spinor representations of $\soKpo$ when $K$ is odd, 
and with type I and type II spinor representations of $\soKpo$ 
when $K$ is even.}

\vs{.1in}
\noindent{\bf Twisted open string partition function of $\suevenK$}
\vs{.1in}

\option
The coefficients 
of the partition function of an open string
stretched between $\omc$-twisted D-branes  $\a$ and $\b$
of $\suevenK$  are given by
\be
\label{eq:suevencoeff}
\nblama 
=  \sum_{ \rrho' = \left\{ {\rm tensors~I} \atop {\rm spinors~I}
		    \right.
            }
{2\; \Spr_{\a'\rrho'} S_{\lam\rrho} \Spr_{\b'\rrho'} 
\over S_{\id\rrho} }  
\ee  
using eqs.~(\ref{eq:ndef}) and (\ref{eq:evenpsi}).
Since the $\omc$-twisted D-branes of $\suevenK$ 
belong to $\Z_2$-cominimal equivalence classes,
we also define the linear combination
\bea
\label{eq:secondsblamadef}
\sblama 
&=& 
\left(\half\right)^{ \half [t(\a)+t(\b)] + 1 }
\left[ {n_{\b\lam}}^{\a}  + {n_{\b\lam}}^{\sig(\a)}  
+ {n_{\sig(\b)\lam}}^{\a}  + {n_{\sig(\b)\lam}}^{\sig(\a)}  \right] 
\nonumber\\
  &=& \left(\half\right)^{ \half [t(\a)+t(\b)] - 2 }
 \sum_{ \rrho' = {\rm tensors~I} }
{ \Spr_{\a'\rrho'} S_{\lam\rrho} \Spr_{\b'\rrho'} 
\over S_{\id\rrho} }  
\eea
where, as a result of eq.~(\ref{eq:soncomin}),
the sum over spinor representations drops out. 
(The normalization is chosen so that $\sblama=\nblama$  
when $\a$ and $\b$ are both type II,
and therefore belong to single-element cominimal equivalence classes.) 
The quantity $\sblama$ is the more natural one
to consider  in the context of level-rank duality.

\section{Twisted states of the $\suoddK$ model} 
\setcounter{equation}{0}
\label{secsuodd}

\vs{.1in}
\noindent{\bf Twisted Ishibashi states of $\suoddK$}
\vs{.1in}

\option
Recall from section \ref{sectwisted} 
that the $\omc$-twisted Ishibashi states $\Ctwishimu$ 
of the $\suoddK$ WZW model
are labelled by self-conjugate integrable highest-weight 
representations $\mu \in \Expc$ of $\suoddK = \AnnoneK$.
The Dynkin indices
$(\mu_0, \mu_1, \cdots, \mu_{n-1}, \mu_n, \mu_n, \mu_{n-1}, \cdots, \mu_1)$
of these representations satisfy
\be
\label{eq:integrableIshiodd}
\mu_0 + 2(\mu_1 + \cdots + \mu_{n})  = K \,.
\ee
Equivalently, the $\omc$-twisted Ishibashi states of $\suoddK$ may be
characterized \cite{Fuchs:1995zr} by the
integrable highest weight representations
of the associated orbit Lie algebra $\gcup = \AnntwoK$,
whose Dynkin diagram is (the right-hand diagram is for $n=1$)

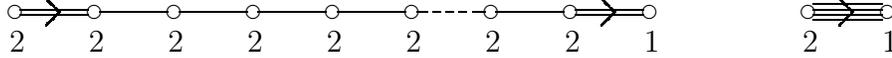
\begin{picture}(500,50)(10,10)
\put(50,40){\circle{5}}
\put(48,25){2}
\put(52,39){\line(1,0){26}}
\drawline(067,40)(062,45)
\drawline(067,40)(062,35)
\put(052,41){\line(1,0){26}}

\put(080,40){\circle{5}}
\put(078,25){2}
\put(082,40){\line(1,0){26}}

\put(110,40){\circle{5}}
\put(108,25){2}
\put(112,40){\line(1,0){26}}

\put(140,40){\circle{5}}
\put(138,25){2}
\put(142,40){\line(1,0){26}}

\put(170,40){\circle{5}}
\put(168,25){2}
\put(172,40){\line(1,0){26}}

\put(200,40){\circle{5}}
\put(198,25){2}
\put(202,40){\line(1,0){4}}
\put(208,40){\line(1,0){3}}
\put(213,40){\line(1,0){3}}
\put(218,40){\line(1,0){3}}
\put(223,40){\line(1,0){5}}

\put(230,40){\circle{5}}
\put(228,25){2}
\put(232,40){\line(1,0){26}}

\put(260,40){\circle{5}}
\put(258,25){2}
\put(262,41){\line(1,0){26}}
\drawline(277,40)(272,45)
\drawline(277,40)(272,35)
\put(262,39){\line(1,0){26}}

\put(290,40){\circle{5}}
\put(288,25){1}

\put(350,40){\circle{5}}
\put(348,25){2}
\put(352,43){\line(1,0){26}}
\put(352,41){\line(1,0){26}}
\drawline(367,40)(362,45)
\drawline(367,40)(362,35)
\put(352,39){\line(1,0){26}}
\put(352,37){\line(1,0){26}}

\put(380,40){\circle{5}}
\put(378,25){1}

\end{picture}

\option
The representation $\mu \in \Expc$ corresponds
to the $\AnntwoK$ representation with Dynkin indices
$(\mu_0, \mu_1,  \cdots,  \mu_n)$.
Consistency with eq.~(\ref{eq:integrableIshiodd})
requires that the dual Coxeter labels be
$(m_0, m_1, \cdots, m_n)=(1,2,2,\cdots,2)$,
and hence we must choose as the \zeroth node
the right-most node of the Dynkin diagrams above
(consistent with ref.~\cite{Gaberdiel:2002qa},
but differing from refs.~\cite{Goddard:1986bp,Fuchsbuch}).

Each $\omc$-twisted Ishibashi state $\mu$ of $\suoddK$
may be mapped \cite{Gaberdiel:2002qa}
to an integrable highest-weight {\it spinor} representation $\mu'$
of the untwisted affine Lie algebra\footnote{
See the note regarding $\sothreeK$ in footnote \ref{fnsothree}.}
$\soKpt$
with Dynkin indices\footnote{
For $n=1$, $\mu'$ has Dynkin indices $(2\mu_0 + 2\mu_1 +3,  2 \mu_1 + 1)$.}
$(\mu_0 + \mu_1 +1, \mu_1, \cdots, \mu_{n-1}, 2 \mu_n + 1)$.
The constraint (\ref{eq:integrableIshiodd}) translates 
into the constraint
$\ell_1 (\mu') \le \smhalf (K+1)$ on the $\soKpt$ Young tableau.
This means that 
{\it $\omc$-twisted Ishibashi states of $\suoddK$ 
are in one-to-one correspondence with the
set of type I spinor representations of $\soKpt$.}

\vs{.1in}
\noindent{\bf Twisted Cardy states of $\suoddK$}
\vs{.1in}

\option
Recall that the $\omc$-twisted Cardy states  $\Ccardya$
(and therefore the $\omc$-twisted D-branes) 
of the $\suoddK$ WZW model
are labelled \cite{Birke:1999ik}
by the integrable highest-weight representations 
$\a \in \Boundomc$ 
of the twisted affine Lie algebra $\twhgK = \AnntwoK$ 
(but see ref.~\cite{Alekseev:2002rj}).
We adopt the same convention as above for the labelling 
of the nodes of the Dynkin diagram of $\AnntwoK$.
Thus the Dynkin indices 
$(a_0, a_1, \cdots, a_n)$ of the highest weights $\a$  
must satisfy
\be
\label{eq:integrableCardyodd}
a_0 + 2 (a_1 + \cdots + a_n) = K\,.
\ee
 
The $\omc$-twisted Cardy state $\a \in \Boundomc$ of $\suoddK$
may be associated \cite{Gaberdiel:2002qa}
 with an integrable highest-weight
{\it spinor} representation $\a'$
of the untwisted affine Lie algebra $\soKpt$
with Dynkin indices\footnote{
For $n=1$, $\a'$ has Dynkin indices $(2 a_0 + 2 a_1 +3,  2 a_1 + 1)$.}
$(a_0+a_1+1, a_1, \cdots, a_{n-1}, 2 a_n + 1 )$.
The constraint (\ref{eq:integrableCardyodd}) translates into the
constraint $\ell_1 (\a') \le \smhalf (K+1)$ on the $\soKpt$ Young tableaux.
This means that 
{\it $\omc$-twisted D-branes of $\suoddK$ 
are in one-to-one correspondence with the
set of type I spinor representations of $\soKpt$.}

Since $\omc$-twisted Cardy states of $\suoddnK$  correspond only to type I 
spinor representations of $\soKpt$,
there is no notion of cominimal equivalence of 
$\omc$-twisted Cardy states in this case.

In the case of $\suoddK$, 
the total number of $\omc$-twisted Cardy states is manifestly equal to 
the total number of $\omc$-twisted Ishibashi states.
The coefficients $\psi_{\a\mu}$  relating 
$\omc$-twisted Cardy states $\a$  to $\omc$-twisted Ishibashi states $\mu$
are given by 
the modular transformation matrix of $\AnntwoK$.
In ref.~\cite{Gaberdiel:2002qa}, it was shown that, for $\suoddK$,
these coefficients are proportional to matrix elements of 
the modular transformation matrix $S'$
of the untwisted affine Lie algebra $\soKpt$:
\be
\label{eq:oddpsi}
\psi_{\a\mu} =   2 S'_{\a'\mu'}
\ee
where $\a'$ and $\mu'$ are the $\soKpt$ representations related
to $\a$ and $\mu$ as described above.

Since the finite Lie algebra associated with
the twisted affine Lie algebra $\AnntwoK$ is $C_n$, 
the representations of $\AnntwoK$ form $C_n$-multiplets at each level.
More specifically \cite{Gaberdiel:2002qa},
each $\omc$-twisted Cardy state $\a \in \Boundomc$ of $\suoddK$
may be associated 
with an integrable highest-weight representation $\a^\dpr$ 
of the untwisted affine Lie algebra $\spnKpn$
with (finite) Dynkin indices 
$(a_1, \cdots, a_n)$.
The row lengths of 
the $\spn$ Young tableau associated with $\a^\dpr$ 
are equal to those of
the $\sooddn$ Young tableau associated with $\a'$ 
reduced by one-half:
$\ell_i(\a'') = \ell_i(\a') - \half$.
Therefore, 
an alternative characterization of 
the $\omc$-twisted D-branes of $\suoddK$
is as the subset of integrable representations of $\spnKpn$
characterized by Young tableaux with row lengths satisfying
$\ell_1 (\a') \le \smhalf K$.

\vs{.1in}
\noindent{\bf Twisted open string partition function of $\suoddK$}
\vs{.1in}

\option
The coefficients 
of the partition function of an open string
stretched between $\omc$-twisted D-branes  $\a$ and $\b$
of $\suoddK$ are given by
\be
\label{eq:suoddcoeff}
\nblama 
=  \sum_{ \rrho' = {\rm spinors~I} }
{ 4\;\Spr_{\a'\rrho'} S_{\lam\rrho} \Spr_{\b'\rrho'} 
\over S_{\id\rrho} }  
\ee  
using eqs.~(\ref{eq:ndef}) and (\ref{eq:oddpsi}).

\vs{.1in}
\noindent{\bf Special case of $\suoddnoddk$}
\vs{.1in}

\option
Note that in the special case of odd level,
the $\omc$-twisted Cardy states $\a$ 
and $\omc$-twisted Ishibashi states $\mu$ of $\suoddnoddk$ 
are in one-to-one correspondence with 
the integrable representations $\a''$ and $\mu''$ of $\spnk$
with finite Dynkin indices
$(a_1, \cdots, a_n)$ and $(\mu_1,  \cdots,  \mu_n)$
respectively.
Moreover, it was observed \cite{Fuchs:1995zr,Petkova:2002yj,Gaberdiel:2002qa}
in this case that the Cardy/Ishibashi coefficients 
may be expressed as 
\be
\label{eq:threegroups}
\psi_{\a\mu} =  S^\dpr_{\a^\dpr \mu^\dpr}  
\ee
where 
$S^\dpr_{\a^\dpr \mu^\dpr}$
are elements of the modular transformation matrix of $\spnk$.

\section{Level-rank duality of the twisted D-branes of $\suNK$}
\setcounter{equation}{0}
\label{secduality}

This section is the heart of the paper, in which
we present the level-rank map between the $\omc$-twisted D-branes
of $\suNK$ and $\suKN$.
We use this to show the level-rank duality of the spectrum of an open
string stretched between $\omc$-twisted D-branes.

As in the case of untwisted D-branes, the level-rank
correspondence involves cominimal equivalence classes
(unless $N$ and $K$ are both odd).
The details of the correspondence differ markedly depending on whether
$N$ and $K$ are even or odd, so we must treat three cases separately.
In refs.~\cite{Naculich:2005tn,Naculich:2006mt},
the tilde ($~\tilde{}~$) notation was used  to denote the level-rank dual
of an untwisted state, 
because the duality map was given by transposition of the associated
Young tableaux.
Here, in all cases, we will use the hat ($~\hat{}~$) notation 
to denote the level-rank dual of an $\omc$-twisted state,
but the specific form of the duality map depends 
on whether $N$ and $K$ are even or odd,
and on whether we are considering $\omc$-twisted Cardy
or $\omc$-twisted Ishibashi states.

\vs{.1in}
\noindent{\bf Duality of twisted states of 
$\suevennevenk \longleftrightarrow \suevenkevenn$}
\vs{.1in}

\option
As we saw in section \ref{secsueven}, 
the cominimal equivalence classes of $\omc$-twisted Cardy states 
(and therefore of $\omc$-twisted D-branes $\a$) of $\suevennevenk$ 
correspond to type I and type II spinor representations $\a'$ of $\sotkpo$.  
The number of equivalence classes 
of $\omc$-twisted D-branes of $\suevennevenk$ is equal to 
the number of equivalence classes 
of $\omc$-twisted D-branes of $\suevenkevenn$,
and there is a natural map $\a \to \ha$ 
between them (when $n$, $k>1$).
This map is defined in terms of the map $\a' \to \ha'$
between the corresponding spinor representations 
of $\sotkpo$ and $\sotnpo$, as follows:
\begin{itemize}
\item
reduce each of the row lengths of $\a'$ by $\smhalf$,
so that they all become integers,
\item
transpose the resulting tableau,
\item
take the complement with respect to an $k \times n$ rectangle,
\item
add $\smhalf$ to each of the row lengths.
\end{itemize}
(The map $\a^\dpr \to \ha^\dpr$
between the corresponding representations of 
${\widehat{\rm sp}(n)_{2k+n-1}}$ and ${\widehat{\rm sp}(k)_{2n+k-1}}$
is given by the middle two steps above.)
The  map $\a' \to \ha'$
was first described in the appendix of ref.~\cite{Mlawer:1990uv} 
in the context of level-rank duality of $\soNK$ WZW models.    
It takes type I (resp.~type II) spinor representations of $\sotkpo$ 
to type II (resp.~type I) spinor representations of $\sotnpo$. 
Hence,  
\be
\label{eq:tplust}
t(\a) + \tildet(\ha) = 1
\ee
for all $\omc$-twisted Cardy states $\a$ of $\suevennevenk$,
where $t(\a)$ is defined in eq.~(\ref{eq:tdef}),
and $\tildet(\ha)$ is the corresponding quantity in $\suevenkevenn$.
The map $\a' \to \ha'$ lifts to a one-to-one map $\a \to \ha$
between cominimal equivalence classes of $\omc$-twisted D-branes 
of $\suevennevenk$ 
and cominimal equivalence classes of $\omc$-twisted D-branes 
of $\suevenkevenn$.

Next, we turn to the level-rank map for $\omc$-twisted
Ishibashi states of $\suevenkevenn$.
As we saw in section \ref{secsueven}, 
$\omc$-twisted Ishibashi states $\mu$ of $\suevennevenk$ 
correspond to  type I tensor and type I spinor representations $\mu'$
of $\sotkpo$.
The level-rank map $\mu \to \hmu$
between $\omc$-twisted Ishibashi states of $\suevennevenk$ 
and those of $\suevenkevenn$
is defined {\it only} for states that correspond 
to type I tensor representations.
The map between $\mu'$ and $\hmu'$,
the corresponding $\sotkpo$ and $\sotnpo$ representations,
is simply given by transposition of the tensor tableaux;
that is, $ \hmu' = \widetilde{ (\mu') } $.
There is no level-rank map between $\omc$-twisted Ishibashi
states that correspond to type I spinor representations,
for the simple reason that these sets of representations 
are not equal in number.  
(Moreover, the map described above for $\omc$-twisted Cardy states
maps type I spinor representations of $\sotkpo$
to type II spinor representations of $\sotnpo$,
which do not correspond to $\omc$-twisted Ishibashi states
of $\suevenkevenn$.)

Having defined the level-rank map between $\mu$ and $\hmu$ 
in terms of the corresponding tensor representations of $\sotkpo$,
one may show that 
\be
\label{eq:firstcominimal}
\hmu = \sig^{-r(\mu)/(2n)} (\tmu) 
\ee
that is, $\hmu$ is in the same $\suevennevenk$ cominimal 
equivalence class (simple-current orbit)
as $\tmu$, where $\tmu$ is the transpose\footnote{
If $\mu$ has $\ell_1 = 2k$, $\tmu$ is obtained by
stripping off leading columns of length $2k$ 
from the transpose of $\mu$.}
of the Young tableau 
of the self-conjugate representation $\mu$ of $\suevennevenk$,
and $r(\mu)$ is the number of boxes of this $\suevennevenk$ tableau.
(Note that $\tmu$ is, in general, not self-conjugate,
while $\hmu$ necessarily is.)
The proof of eq.~(\ref{eq:firstcominimal}) 
is very similar to one given in section 6 of ref.~\cite{Naculich:2006mt}.
A consequence of eq.~(\ref{eq:firstcominimal}) is that 
the $\suevennevenk$ modular transformation matrix $S$
is related to 
the $\suevenkevenn$ modular transformation matrix $\tS$
by
\be
\label{eq:firstsimple}
S^*_{\lam\mu} = \sqrt{k \over n} \tS_{\tlam\hmu}
\ee
which follows from \cite{Altschuler:1989nm,Mlawer:1990uv}
\bea
S^*_{\lam\mu} 
&=& \sqrt{k \over n} \e^{ 2\pi i r(\lam) r(\mu)/(4nk)} \tS_{\tlam\tmu} \,,
\nonumber\\
\tS_{\tlam\tmu} 
&=& 
\e^{-2\pi i r(\lam) r(\mu)/(4nk)}  \tS_{\tlam\hmu}\,.
\eea

Having defined level-rank maps for the $\omc$-twisted Cardy and
Ishibashi states of $\suevennevenk$,
we now turn to the duality of the open-string spectrum
between $\omc$-twisted D-branes.
The coefficients of the partition function 
of an open string stretched between $\omc$-twisted D-branes $\a$ and $\b$
are real numbers so we may write (\ref{eq:secondsblamadef}) as
\be
\sblama 
= \left(\half\right)^{ \half [t(\a)+t(\b)] - 2 }
 \sum_{ \rrho' = {\rm tensors~I} }
{ \Spr_{\a'\rrho'} S^*_{\lam\rrho} \Spr_{\b'\rrho'} 
\over S^*_{\id\rrho} }  \,.
\ee
In ref.~\cite{Mlawer:1990uv}, 
the spinor-tensor components $S'_{\a'\mu'}$ 
of the modular transformation matrix of $\sotkpo$ 
were shown to be related to 
the spinor-tensor components $\tS'_{\ha'\hmu'}$ of $\sotnpo$ by
\be
\label{eq:tensorspinor}
S'_{\a'\mu'}  = 
2^{t(\a) -{1\over 2}} (-1)^{r(\mu')}
  \tS'_{\ha'\hmu'}
= 2^{{1\over 2} [t(\a) -\tildet(\ha)]} (-1)^{r(\mu')}
  \tS'_{\ha'\hmu'}
\ee
where we have used eq.~(\ref{eq:tplust}).
Using eqs.~(\ref{eq:firstsimple}) and (\ref{eq:tensorspinor}),
we find 
\be
\sblama 
= \left(\half \right)^{ \half [\tildet(\ha)+\tildet(\hb)]  -2 }
 \sum_{ \hrho' = {\rm tensors~I} }
{ \tSpr_{\ha'\hrho'}  \tS_{\tlam\hrho} \tSpr_{\hb'\hrho'} 
           \over \tS_{\id\hrho} }  
= \tsblama \,.
\ee
Thus the (linear combination of) coefficients 
(\ref{eq:secondsblamadef})
of the open-string partition function
of $\omc$-twisted D-branes 
of $\suevennevenk$
are equal to those of $\suevenkevenn$
under the level-rank duality map acting on $\omc$-twisted D-branes.

\vs{.1in}
\noindent{\bf Duality of twisted states of 
$\suoddnoddk \longleftrightarrow \suoddkoddn$}
\vs{.1in}

\option
As we saw in section \ref{secsuodd},
the $\omc$-twisted Cardy states 
(and therefore the $\omc$-twisted D-branes $\a$) of $\suoddnoddk$
map one-to-one to type I spinor integrable representations $\a'$ 
of $\sotkpth$,
and also to integrable representations $\a''$ of $\spnk$.
We define the level-rank duality map 
$\a \to \ha$ for $\omc$-twisted Cardy states 
by transposition of the associated $\spnk$ tableaux:  
that is, $\ha^\dpr = {\widetilde{ (\a^\dpr) }} $.
(In ref.~\cite{Naculich:2006mt}, we therefore denoted this map simply
by $\a \to \ta$.)
Exactly similar statements hold for the 
$\omc$-twisted Ishibashi states $\mu$ of $\suoddnoddk$.

The equality of the Cardy/Ishibashi coefficients 
of $\suoddnoddk$ and $\suoddkoddn$
\be
\psi_{\a\mu} =    \tpsi_{\ha\hmu}
\ee
follows immediately from eq.~(\ref{eq:threegroups})
together with level-rank duality of the $\spnk$ WZW model \cite{Mlawer:1990uv} 
\be
\label{eq:spnSduality}
S^\dpr_{\a^\dpr\mu^\dpr} = \tS^\dpr_{\ha^\dpr\hmu^\dpr}
\ee
where $S^\dpr$ and $\tS^\dpr$ are the modular transformation 
matrices of $\spnk$ and $\spkn$ respectively.
Moreover, by eq.~(\ref{eq:oddpsi}), we have
\be
\label{eq:sotkpthduality}
S'_{\a'\mu'}  = \tS'_{\ha'\hmu'}
\ee
where $S'$ and $\tS'$ are the modular transformation matrices of
$\sotkpth$ and $\sotnpth$ respectively,
and the map $\a' \to \ha'$ 
from $\sotkpth$ to $\sotnpth$ 
(induced from the transposition map $\a^\dpr \to \ha^\dpr$)
is:

\begin{itemize}

\item  reduce each of the row lengths of $\a'$ by $\smhalf$,
so that they all become integers,
\item transpose the resulting tableau,
\item add $\smhalf$ to each of the row lengths.

\end{itemize}
\noindent 
and  equivalently for $\mu' \to \hmu'$.
(Note that this map differs from spinor map defined in the
last subsection by the omission of the complement map.)
Note that eq.~(\ref{eq:sotkpthduality}) differs from
the standard level-rank duality of WZW models \cite{Mlawer:1990uv},
which relates $\soNK$ to $\soKN$.

Finally, 
we turn to the duality of the open-string spectrum
between $\omc$-twisted D-branes of $\suoddnoddk$ and $\suoddkoddn$.
In ref.~\cite{Naculich:2006mt},  
$\spnk$ level-rank duality (\ref{eq:spnSduality})
was used to show the level-rank duality of the
coefficients of the open string partition function.
We can equivalently use 
eqs.~(\ref{eq:suoddcoeff}) and (\ref{eq:sotkpthduality}) 
to show the same result
\be
\nblama 
=  \sum_{ \rrho' = {\rm spinors~I}  }
{4\; \Spr_{\a'\rrho'} S^*_{\lam\rrho} \Spr_{\b'\rrho'} 
\over S^*_{\id\rrho} }  
=  \sum_{ \hrho' = {\rm spinors~I}  }
{4\; \tSpr_{\ha'\hrho'} \tS_{\tlam\hrho} \tSpr_{\hb'\hrho'} 
\over \tS_{\id\hrho} } 
=\tnblama
\ee
since $\mu' \to \hmu'$ maps 
type I spinor representations of $\sotkpth$ to 
type I spinors of $\sotnpth$,
and we have also used 
\be
\label{eq:secondsimple}
S^*_{\lam\mu} = \sqrt{2k+1 \over 2n+1} \tS_{\tlam\hmu}
\ee
which was proved in ref.~\cite{Naculich:2006mt}.

\vs{.1in}
\noindent{\bf Duality of twisted states of 
$\suoddnevenk \longleftrightarrow \suevenkoddn$}
\vs{.1in}

\option
Recall that the $\omc$-twisted D-branes $\a$ of $\suoddnevenk$ 
correspond to  
type I spinor representations $\a'$ of $\sotkpt$,  
and the equivalence classes 
of $\omc$-twisted D-branes $\ha$ of $\suevenkoddn$ 
correspond to  
type I spinor representations $\ha'$ of $\sotnpt$.  
The number of such spinor representations is equal,
and we define the one-to-one level-rank map $\a' \to \ha'$ 
from $\sotkpt$ to $\sotnpt$  
(for $k>1$) as follows:

\begin{itemize}

\item  reduce each of the row lengths of $\a'$ by $\smhalf$,
so that they all become integers,
\item transpose the resulting tableau,
\item take the complement with respect to an $k \times n$ rectangle,
\item add $\smhalf$ to each of the row lengths.

\end{itemize}
\option
(By comparison, 
the definition of $\a' \to \ha'$ from $\sotkpo$ to $\sotnpo$ is the same,
but in that case type I spinors are mapped to type II spinors and vice versa.)
The map $\a' \to \ha'$ 
lifts to a one-to-one map $\a \to \ha$ between
$\omc$-twisted D-branes of $\suoddnevenk$
and {\it equivalence classes of}
$\omc$-twisted D-branes of $\suevenkoddn$.
(The map $\a^\dpr \to \ha^\dpr$
between the corresponding representations of 
${\widehat{\rm sp}(n)_{2k+n}}$ and ${\widehat{\rm sp}(k)_{2n+k}}$
is given by the middle two steps above.)

Next, we turn to the level-rank map between $\omc$-twisted
Ishibashi states.
The $\omc$-twisted Ishibashi states $\mu$ of $\suoddnevenk$ 
correspond to  type I spinor representations $\mu'$ of $\sotkpt$.  
The $\omc$-twisted Ishibashi states $\hmu$  of $\suevenkoddn$ 
correspond to type I tensor and type I spinor representations  $\hmu'$
of $\sotnpt$.  
The number of such representations on each side is not equal,
and the level-rank map $\mu' \to \hmu'$ takes type I spinor
representations of $\sotkpt$ to {\it only} the type I tensor
representations of $\sotnpt$.
(Just as for $\suevenkevenn$, 
there is no level-rank correspondence 
for the spinor Ishibashi states of $\suevenkoddn$.)
The map $\mu' \to \hmu'$ from $\sotkpt$ to $\sotnpt$  
is defined as follows:

\begin{itemize}
\item reduce each of the row lengths of $\mu'$ by $\smhalf$,
so that they all become integers, and
\item transpose the resulting tableau.
\end{itemize}

The map $\mu' \to \hmu'$ then lifts to a map $\mu \to \hmu$ between 
$\omc$-twisted Ishibashi states of $\suoddnevenk$ 
and a subset of $\omc$-twisted Ishibashi states of $\suevenkoddn$.
One may show that 
\be
\label{eq:thirdcominimal}
\hmu = \sig^{-r(\mu)/(2n+1)} (\tmu) 
\ee
where $\tmu$ is the transpose\footnote{
If $\mu$ has $\ell_1 = 2k$, $\tmu$ is obtained by
stripping off leading columns of length $2k$ 
from the transpose of $\mu$.}
of the Young tableau 
of the self-conjugate representation $\mu$ of $\suoddnevenk$,
and $r(\mu)$ is the number of boxes of this $\suoddnevenk$ tableau.
The proof of eq.~(\ref{eq:thirdcominimal}) is very similar to 
one given in section 6 of ref.~\cite{Naculich:2006mt}.
Consequently, the modular transformation matrices
$S$ of $\suoddnevenk$ and 
$\tS$ of $\suevenkoddn$ are related by 
\be
\label{eq:thirdsimple}
S^*_{\lam\mu} = \sqrt{2k \over 2n+1} \tS_{\tlam\hmu}
\ee
which follows from \cite{Altschuler:1989nm,Mlawer:1990uv}
\bea
S^*_{\lam\mu} 
&=& \sqrt{2k \over 2n+1} \e^{ 2\pi i r(\lam) r(\mu)/(2n+1)(2k)} \tS_{\tlam\tmu} \,,
\nonumber\\
\tS_{\tlam\tmu} 
&=& 
\e^{-2\pi i r(\lam) r(\mu)/(2n+1)(2k) }  \tS_{\tlam\hmu} \,.
\eea

Finally, in the appendix of this paper, we show that
\be
\label{eq:sotkptduality}
S'_{\a'\mu'} = (-1)^{r(\hmu')+k}  \tS'_{\ha'\hmu'}
\ee
where $S'_{\a'\mu'}$  and $\tS'_{\ha'\hmu'}$ 
are modular transformation matrices of $\sotkpt$ and $\sotnpt$ respectively.
As before,
we observe that eq.~(\ref{eq:sotkptduality})
is {\it not} the standard $\soNK \leftrightarrow \soKN$ duality of WZW models.

Equations (\ref{eq:thirdsimple}) and (\ref{eq:sotkptduality}) 
may be used to establish
the level-rank duality of the coefficients of the partition 
functions (\ref{eq:suoddcoeff}) and (\ref{eq:secondsblamadef})
of an open string stretched between $\omc$-twisted D-branes 
of $\suoddnevenk$ and $\suevenkoddn$
\be
\nblama 
=  \sum_{ \rrho' = {\rm spinors~I} }
{4\; \Spr_{\a'\rrho'} S^*_{\lam\rrho} \Spr_{\b'\rrho'} 
\over S^*_{\id\rrho} } 
 =\sum_{ \hrho' = {\rm tensors~I} }
{4\; \tSpr_{\ha'\hrho'} \tS_{\tlam\hrho} \tSpr_{\hb'\hrho'} 
\over \tS_{\id\hrho} }
= \tsblama
\ee
where the last equality follows because $\ha$ and $\hb$ are both type I 
spinor representations of $\sotnpt$,
so that $\tildet(\ha)=\tildet(\hb)=0$.

\section{Level-rank duality of twisted D-brane charges}
\setcounter{equation}{0}
\label{seccharge}

In this section, we ascertain the relationship between the charges
of level-rank-dual $\omc$-twisted D-branes of $\suNK$ and $\suKN$.
Recall from ref.~\cite{Gaberdiel:2003kv} 
that the D0-brane charge of the $\om_c$-twisted D-brane of $\suNK$ 
labelled by $\a$ is given by 
\be
\label{eq:twcharge}
Q^\omc_\a = (\dim \a^\dpr)_\spn  \quad 
\mod \xNK \qquad {\rm for}~~ \suNK 
\ee
where $\a^\dpr$ is the $\spn$ representation corresponding to the
$\omc$-twisted Cardy state $\a$ of $\suevenK$ or $\suoddK$,
as described in sections~\ref{secsueven} and \ref{secsuodd}.

Since the charges of $\suNK$ D-branes (both untwisted and twisted)
are defined only modulo $\xNK$,
and those of $\suKN$ D-branes modulo $\xKN$,
comparison of charges of level-rank-dual D-branes 
is only possible modulo 
$x \equiv \gcd \{ \xNK, \xKN  \} = \min \{ \xNK, \xKN  \}$.
In refs.~\cite{Naculich:2005tn,Naculich:2006mt}, 
the charges of untwisted D-branes of the $\suNK$ model 
and those of the level-rank-dual $\suKN$ model 
were shown to be equal modulo $x$, up to a (known) sign
(\ref{eq:untwistedduality}), (\ref{eq:specialduality}).
In ref.~\cite{Naculich:2006mt}, 
the charges of $\omc$-twisted D-branes of the $\suoddnoddk$ model 
and those of the level-rank-dual $\suoddkoddn$ model 
were also shown to be equal, modulo $x$.
As we will see below, the relationship between charges
of level-rank-dual $\omc$-twisted D-branes of $\suNK$ and $\suKN$
is more complicated 
when $N$ and $K$ are not both odd.

\vs{.1in}
\noindent{\bf Charges of cominimally-equivalent twisted D-branes
of $\suevenK$}
\vs{.1in}

\option
Since level-rank duality is a correspondence between $\Z_2$-cominimal 
equivalence classes of $\omc$-twisted D-branes when either $N$ or $K$
is even, 
we must first demonstrate that
cominimally-equivalent $\omc$-twisted D-branes of $\suevenK$ 
have the same charge 
(modulo sign and modulo $\xevennK$).
The $\spn$ representation  $\a^\dpr$ is related to 
the $\sooddn$ representation $\a'$
by reducing each row length of the tableau for the latter by one-half.  
As demonstrated in appendix A of ref.~\cite{Gaberdiel:2003kv} 
(see also ref.~\cite{Bouwknegt:2006pd}), 
the respective dimensions of these representation are related by 
the ``miraculous dimension formula''
\be
\label{eq:mirac}
(\dim \a')_\sooddn  = 2^n~ (\dim \a^\dpr)_\spn  \,. 
\ee
Next, in appendix B of ref.~\cite{Gaberdiel:2003kv},
it is shown that
\be
\label{eq:appb}
(\dim \sig(\lam))_\sooddn  = ~-~ (\dim \lam)_\sooddn  \quad \mod \xevennK
\ee
where $\sig(\lam)$ is the $\soKpo$  representation 
cominimally-equivalent to $\lam$.
Using conjecture B${}^{\rm spin}$ of ref.~\cite{Gaberdiel:2003kv},
and the facts that the dimensions of all spinor representations 
of $\sooddn$ are multiples of $2^n$
and that 
$(\dim \sig(0))_\sooddn  = -1 ~\mod \xevennK$ \cite{Gaberdiel:2003kv},
eq.~(\ref{eq:appb}) may be strengthened to 
\be
(\dim \sig(\a'))_\sooddn  = ~-~ (\dim \a')_\sooddn  \quad \mod 2^n ~\xevennK
\ee
for $\a'$ a spinor representation of $\soKpo$. 
Together with eq.~(\ref{eq:mirac}), this implies that 
the charges of cominimally-equivalent $\omc$-twisted D-branes 
of $\suevenK$ are related by
\be
Q^\omc_{\sig(\a)} = -Q^\omc_\a \quad \mod \xevennK
\ee
analogous to eq.~(\ref{eq:untwistedcomin}) for untwisted D-branes.

Finally, we turn to the relationship between the charges of
level-rank-dual $\omc$-twisted D-branes.

\vs{.1in}
\noindent{\bf Duality of twisted D-brane charges under 
$\suoddnoddk \longleftrightarrow \suoddkoddn$}
\vs{.1in}

\option
Let  $x = \gcd \{ \xoddnoddk, \xoddkoddn  \} $.
In ref.~\cite{Naculich:2006mt}, it was shown that 
\be
(\dim \a^\dpr)_\spn  = (\dim \ha^\dpr)_\spk \quad \mod x
\ee
where  $\ha^\dpr$ is obtained from $\a^\dpr$ by tableau transposition.
Since $\ha^\dpr$ is the $\spk$ representation corresponding to the
level-rank-dual $\omc$-twisted D-brane $\ha$ of $\suoddkoddn$,
it immediately follows from eq.~(\ref{eq:twcharge}) that the charges
of level-rank-dual $\omc$-twisted D-branes are  equal
\be
Q^\omc_\a = \tQ^\omc_\ha \quad \mod x  \,.
\ee
This was previously presented in ref.~\cite{Naculich:2006mt}
and is included here for completeness.

\vs{.1in}
\noindent{\bf Duality of twisted D-brane charges under 
$\suoddnevenk \longleftrightarrow \suevenkoddn$}
\vs{.1in}

\option
Let  $x = \gcd \{ \xoddnevenk, \xevenkoddn  \} $.
We begin with the relationship
\be
(\dim \Lam_s)_{\spn} 
= (\dim \Lam_s)_{\suoddn} -(\dim \Lam_{s-1})_{\suoddn} 
\ee
where $\Lam_s$ is the completely antisymmetric representation
with Young tableau $ \oneoneoneone \} s $.
Next, 
as shown in ref.~\cite{Naculich:2005tn},
\be
(\dim \Lam_s)_{\suoddn} = (-1)^s (\dim \tLam_{s})_{\suevenk} 
\quad \mod x
\ee 
where $\tLam_s$ is the completely symmetric representation
with Young tableau $\underbrace{\four}_s$.
Finally, 
\be
(\dim \tLam_s)_{\sooddk} 
= (\dim \tLam_s)_{\suevenk} +(\dim \tLam_{s-1})_{\suevenk} \,.
\ee
Combining these three equations, we obtain
\be
(\dim \Lam_s)_{\spn} = (-1)^s (\dim \tLam_s)_{\sooddk} \quad \mod x\,.
\ee
This result can be used 
in the determinantal formulas (A.44) and (A.60) of ref.~\cite{Fultonbook},
following the approach of ref.~\cite{Naculich:2006mt},
to establish a relationship between arbitrary representations 
of $\spn$ and $\sooddk$,
\be
\label{eq:arbit}
(\dim \a^\dpr)_{\spn} 
= (-1)^{r(\a^\dpr)} (\dim \tadp )_{\sooddk} \quad \mod x
\ee
where $\tadp$  is the transpose of the tableau of $\a^\dpr$.

Now, from the level-rank map of section~\ref{secduality}, 
the representation $\tadp$ is related to the
representation $\ha'$ 
that corresponds to the level-rank dual $\omc$-twisted D-brane
by taking the complement of the tableau
with respect to a $k \times (n + \half)$ rectangle.
This maps a type I tensor representation of
$\sotnpt$ to a type I spinor representation.
We conjecture a relationship\footnote{
After v1 of this paper appeared, 
we learned that an equivalent version of this relationship 
has been independently conjectured by 
Stefan Fredenhagen and collaborators \cite{private}.}
\be
\label{eq:unexpected}
(\dim \tadp )_{\sooddk} 
= (-1)^{k(k+1)/2} (\dim \ha')_{\sooddk} 
\quad \mod \xevenkoddn     \qquad{\rm (conjecture)}
\ee
between the dimensions of $\tadp$ and   $\ha'$.
To justify this, consider the expression for the dimension of 
the $\sooddk$ representation $ \tadp$:
\be
\label{eq:firstrep}
(\dim \tadp )_{\sooddk} 
=  { \prod_{i=1}^k  (2 \phi_i)
  \prod_{i<j}  (\phi_i - \phi_j) (\phi_i + \phi_j)
	\over
   \prod_{i=1}^k  (2k + 1 - i)
  \prod_{i<j}  (j-i) (2k+1-i-j) }
\ee
where $\phi_i = \ell_i (\tadp) - \half + k - i$.
All the factors in parentheses are integers.
The row lengths of $\ha'$ are related to those of $\tadp$ by
$\ell_i (\ha') =  n+\half- \ell_{k+1-i} (\tadp)$. 
Hence 
\be
\label{eq:secondrep}
(\dim \ha')_{\sooddk} 
= { \prod_{i=1}^k  (X - 2 \phi_{k+1-i} )
  \prod_{i<j}  
  (\phi_{k+1-j} - \phi_{k+1-i}) (X - \phi_{k+1-i} - \phi_{k+1-j})
  \over
   \prod_{i=1}^k  (2k + 1 - i)
  \prod_{i<j}  (j-i) (2k+1-i-j) }
\ee
where $X \equiv 2n + 2k +1$.
Then
\be
\label{eq:diff}
(\dim \tadp )_{\sooddk} 
- (-1)^{k(k+1)/2}
(\dim \ha')_{\sooddk} 
=  X R
\ee
where $R$ is a rational number with denominator
${ \prod_{i=1}^k  (2k + 1 - i) \prod_{i<j}  (j-i) (2k+1-i-j) }$.
If $X$ is prime, then none of the factors in the denominator of $R$
(which are all less than $2k+1$) divide $X$, 
and since the left-hand-side is an integer, $R$ must also be an integer, 
in which case the left-hand-side is a multiple
of $X$.     
This establishes eq.~(\ref{eq:unexpected}) when $X$ is prime,
since $\xevenkoddn = X$ in that case.
When $X$ is not prime, some of the factors in the denominator
of $R$ may divide $X$, but we believe 
(proved for $k=2$, and based on strong numerical evidence 
for $k=3$, 4, and 5, with arbitrary $n$)
that the right-hand-side of
eq.~(\ref{eq:diff}) is always a multiple of $\xevenkoddn$,
and therefore that the conjecture (\ref{eq:unexpected}) holds.

Finally, from  eq.~(\ref{eq:mirac}), we have
\be 
\label{eq:mirack}
(\dim \ha')_{\sooddk}  =  2^k (\dim \ha^\dpr)_{\spk}  \,.
\ee
Putting together 
eqs.~(\ref{eq:arbit}), (\ref{eq:unexpected}), and (\ref{eq:mirack}), 
we obtain the relationship between the charge of
the $\omc$-twisted  D-brane $\a$ of $\suoddnevenk$
and the level-rank-dual $\omc$-twisted D-brane $\ha $
of $\suevenkoddn$
\be
\label{eq:complic}
Q^\omc_\a =  2^k\, 
(-1)^{r(\a^\dpr)+k(k+1)/2}  \,\tQ^\omc_\ha \quad \mod x 
\ee
whose validity is subject only to the conjectured relation
(\ref{eq:unexpected}).

\vs{.1in}
\noindent{\bf Duality of twisted D-brane charges under 
$\suevennevenk \longleftrightarrow \suevenkevenn$}
\vs{.1in}

\option
Let  $x = \gcd \{ \xevennevenk, \xevenkevenn  \} $.
As shown in ref.~\cite{Naculich:2006mt},
if $n=k$, then $x = 4$ if $n= 2^m$, otherwise $x=1$.
If $n\neq k$, then $x = 2$ if $n+k = 2^m$,  otherwise $x=1$.

We saw above that the charges of 
level-rank dual $\omc$-twisted D-branes of $\suNK$ and $\suKN$
are equal (modulo $x$) when both $N$ and $K$ are odd.
This equality (modulo $x$) no longer holds if either $N$ or $K$ is even.
When both $N$ and $K$ are even,  
the charges are again not equal (even modulo $x$ and modulo sign),
as may be checked in a specific case 
(e.g., 
${\widehat{\rm su}(4)_4}$, with $\a^\dpr=\oneone$ and $\ha^\dpr=\two$,
since $ 5 \neq \pm 10$ mod 4).
On the basis of eq.~(\ref{eq:complic}),
one might expect a relationship such as 
\be
2^n Q^\omc_\a = 
\pm  2^k \tQ^\omc_\ha \quad \mod x  \,.
\ee
However,
any such relationship is trivially satisfied,
since $\omc$-twisted branes exist only when $n$, $k > 1$,
and $x$ is either 1, 2, or 4.

\section{Conclusions}
\setcounter{equation}{0}
\label{secconcl}

In this paper, we have
considered D-branes of the $\suNK$ WZW model
twisted by the charge-conjugation symmetry $\omc$.
Such D-branes exist for all $N>2$,
and possess integer D0-brane charge, defined modulo $\xNK$.

For $\suevenK$ and $\suoddK$, 
the $\omc$-twisted D-branes are labelled by 
a subset of the integrable representations 
of $\soKpo$ and $\soKpt$ respectively.
In the former case, the D-branes belong to
cominimal equivalence classes generated by the
$\Z_2$ simple current symmetry of $\soKpo$.
We showed that the D0-brane charges of cominimally equivalent D-branes
are equal and opposite modulo $\xevennK$.

We then showed that level-rank-duality of $\suNK$ WZW models
extends to the $\omc$-twisted D-branes of the theory
when both $N$ and $K$ are greater than two.
In particular, 
we demonstrated a one-to-one mapping $\a \to \ha$ 
between the
$\omc$-twisted D-branes for $N$ odd 
(or cominimal equivalence classes of D-branes for $N$ even)
of $\suNK$ and 
the $\omc$-twisted D-branes for $K$ odd 
(or cominimal equivalence classes of D-branes for $K$ even)
of $\suKN$.

We then showed that the spectrum of an open string stretched between 
$\omc$-twisted D-branes is invariant under level-rank duality. 
More precisely, we showed that 
the coefficients $\nblama$ of the open-string partition function 
(or $\sblama$,  the appropriate linear combination (\ref{eq:firstsblamadef})
of those coefficients corresponding to cominimal equivalence classes
of $\omc$-twisted D-branes of $\suevenK$)
are invariant under $\a \to \ha$, $\b \to \hb$, and $\lam \to \tlam$.
The proof of this required the existence of a {\it partial} 
level-rank mapping between the $\omc$-twisted Ishibashi states 
of each theory.   (That is, the map only involved a subset of 
the $\omc$-twisted Ishibashi states of $\suevenK$.)

Finally, we analyzed the relation between the D0-brane charges
of level-rank-dual $\omc$-twisted D-branes (or cominimal
equivalence classes thereof), modulo $x = \gcd\{\xNK, \xKN\}$. 
When $N$ and $K$ are both odd, the charges are equal mod $x$
(as previously demonstrated in ref.~\cite{Naculich:2006mt}),
but in other cases this simple relationship does not hold.
For $N = 2n+1$ and $K= 2k$, the relation between the charges 
of level-rank-dual $\omc$-twisted D-branes is
\be
Q^\omc_\a =  2^k\, 
(-1)^{r(\a^\dpr)+k(k+1)/2}  \,\tQ^\omc_\ha \quad \mod x 
\ee
subject to the validity of a certain conjecture (\ref{eq:unexpected}) stated 
in section~\ref{seccharge}.

It would interesting to know whether level-rank duality
extends to any of the other twisted D-branes of the $\suNK$ WZW
model \cite{Gaberdiel:2004hs}.

\section*{Acknowledgments}

The authors wish to express their gratitude 
to M.~Gaberdiel for several extremely helpful 
correspondences.
We also thank P.~Bouwknegt and D.~Ridout for 
useful comments.

\section*{Appendix}
\setcounter{equation}{0}
\def\theequation{A.\arabic{equation}}

In this appendix, we establish the relationship
between certain matrix elements
of the modular transformation matrices
of $\sotkpt$ and $\sotnpt$ through the use of
Jacobi's theorem, following the approach of ref.~\cite{Mlawer:1990uv}.
Note that this is {\it not} the usual level-rank duality 
between $\soNK$ and $\soKN$.

Let $\a'$ ($\mu'$) be an integrable type I spinor representation 
of $\sotkpt$ corresponding to a $\omc$-twisted Cardy (Ishibashi)
state of $\suoddnevenk$.
The $\sotkpt$ modular transformation matrix 
has the matrix element \cite{Kac:1984mq} 
\be
\label{eq:appfirst}
S'_{\a'\mu'} 
 =  (-1)^{n(n-1)/2}  2^{n-1} (2k + 2n + 1)^{-n/2} \det \bM    
\ee
where $\bM$ is an $n \times n$  matrix with matrix elements
\be
M_{ij}  =  \sin 
\left( \pi \phi_i(\a') \phi_j(\mu') \over k + n + \smhalf \right) ,
\qquad
\phi_i (\a') 
= \ell_i(\a') + n + \smhalf - i, \qquad i=1,\cdots, n\,.
\ee
Let $\ha'$ ($\hmu'$) be the integrable type I spinor (tensor) representation
of $\sotnpt$ related to $\a'$ ($\mu'$) by the level-rank duality map
described in section \ref{secduality}.  
The $\sotnpt$ modular transformation matrix 
has matrix element
\be
\label{eq:appsecond}
\tS'_{\ha'\hmu'} 
 =  (-1)^{k(k-1)/2}  2^{k-1} (2k + 2n + 1)^{-k/2} \det  \tbM
\ee
where $\tbM$ is a $k \times k$ matrix with matrix elements
\be
\tM_{ij}  = \sin 
\left( \pi \tphi_i(\ha') \tphi_j(\hmu') \over k + n + \smhalf \right) ,
\qquad
\tphi_i (\ha') 
= \ell_i(\ha') + k + \smhalf - i, \qquad i=1,\cdots, k \,.
\ee
Next, define the index sets for the $\omc$-twisted Cardy states
\bea
I & = & \{ \phi_i (\a'),  \qquad i=1, \cdots, n \}, \nonumber\\
\bI & = & \{ \tphi_i (\ha'), \qquad  i=1, \cdots, k \}  \,.
\eea
Using the level-rank duality map $\a' \to \ha'$ 
given in section~\ref{secduality},
one may establish that $I$ and $\bI$ are complementary sets of integers:
\be
I \cup \bI = \{1, 2, \cdots, n+k \}, \qquad I \cap \bI = 0\,.
\ee
Also, define the index sets for the $\omc$-twisted Ishibashi states
\bea
J  &=&  \{ \phi_j (\mu'), \quad j=1, \cdots, n \}, \nonumber\\
\bJ  &=&  \{ n + k + \smhalf - \tphi_j (\hmu'), \quad j=1, \cdots, k \}.
\eea
Using the level-rank duality map $\mu' \to \hmu'$ 
given in section~\ref{secduality},
one may also establish that $J$ and $\bJ$ are complementary sets of integers:
\be
J \cup \bJ = \{1, 2, \cdots, n+k \}, \qquad J \cap \bJ = 0\,.
\ee
Now, define the $L \times L$ matrix $\bOm$ with matrix elements
\be
\Omega_{ij} = \sin \left( \pi ij \over {L+\smhalf} \right), 
\qquad i, j= 1, \cdots, L
\ee
where $L = n+k$.  This matrix has determinant
\be
\label{eq:appthird}
\det \bOm = (-1)^{L(L-1)/2} \left( 2L + 1 \over 4 \right)^{L/2} 
\ee
and obeys 
\be
\label{eq:appfourth}
\bOm^{-1} = 
\left(4 \over  2L + 1 \right) \bOm\,.
\ee
Define $(\bOm)_{IJ}$
to be the $n \times n$ submatrix obtained from the larger $\bOm$ 
by considering only those rows indexed by the elements of $I$
and those columns indexed by the elements of $J$.  
Jacobi's theorem \cite{Jeffreysbook} states that
\be
\label{eq:appfifth}
\det [(\bOm^{-1})^T]_{IJ} = (-)^{ \Sigma_I + \Sigma_J} (\det \bOm)^{-1}
 \det (\bOm)_{\bI\,\bJ}~,
\ee
where
\be
\Sigma_I = \sum_{i \in I}i
{\rm~~~and~~~}
\Sigma_J = \sum_{j \in J}j~.
\ee 
One may observe that 
\be 
\label{eq:appsixth}
\det \bM = \det (\bOm)_{IJ}, \qquad
\det \tbM = (-1)^{k + \Sigma_{\bI} + k(k-1)/2} \det  (\bOm)_{\bI \bJ}
\ee
where the last contribution to the sign results from reversing
the ordering of the rows of $\tbM$.
Assembling eqs.~(\ref{eq:appfirst}),
(\ref{eq:appsecond}),
(\ref{eq:appthird}),
(\ref{eq:appfourth}),
(\ref{eq:appfifth}), and
(\ref{eq:appsixth}),
and using
\be
(-1)^{\Sigma_I + \Sigma_J + \Sigma_{\bI}}
= (-1)^{\Sigma_{\bJ}} = (-1)^{nk + k(k-1)/2 + r(\hmu')}
\ee
one concludes that
\be
S'_{\a'\mu'} = (-1)^{r(\hmu')+k}  \tS'_{\ha'\hmu'}
\ee
which is used in proving the level-rank duality of the
open string spectrum in the last subsection of section \ref{secduality}.

\providecommand{\href}[2]{#2}\begingroup\raggedright\endgroup

\end{document}